\documentclass[10pt,a4paper]{article}
\usepackage{amsmath,amssymb,amsthm}

\newtheorem{theorem}{Theorem}
\newtheorem{corollary}{Corollary}

\newcommand{\mf}{\mathfrak}
\newcommand{\mc}{\mathcal}

\newcommand{\Complex}{\mathbb{C}}

\newcommand{\Natural}{\mathbb{N}}

\title{Branching rules of semi-simple Lie algebras\\using affine extensions}
\date{February 2002}
\author{T. Quella\footnote{E-mail: quella@aei-potsdam.mpg.de}\\\\
  Max-Planck-Institut f\"ur Gravitationsphysik\\(Albert-Einstein-Institut)\\
  Am M\"uhlenberg 1\\D-14476 Golm\\ Germany}

\begin{document}

\maketitle
\vspace{-9cm}\hfill AEI-2001-133, math-ph/0111020\vspace{8cm}

\begin{abstract}
  We present a closed formula for the branching coefficients of an embedding
  $\mf{p}\hookrightarrow\mf{g}$ of two
  finite-dimensional semi-simple Lie algebras. The formula
  is based on the untwisted affine
  extension of~$\mf{p}$. It leads to an alternative proof
  of a simple algorithm for the computation of branching rules which
  is an analog of the Racah-Speiser algorithm for tensor products.
  We present some simple applications and describe how
  integral representations for branching coefficients can be obtained.
  In the last part we comment on the relation of our approach
  to the theory of NIM-reps of the fusion ring in
  WZW models with chiral algebra~$\hat{\mf{g}}_k$. In fact, it turns out
  that for these models each embedding $\mf{p}\hookrightarrow\mf{g}$
  induces a NIM-rep at level $k\to\infty$. In cases where
  these NIM-reps can be extended to finite level, we obtain a
  Verlinde-like formula for branching coefficients. Reviewing this question
  we propose a solution to a puzzle which remained open in related
  work by Alekseev, Fredenhagen, Quella and Schomerus.
\end{abstract}

  \centerline{PACS: 02.20.Sv, 11.25.Hf\hspace{3cm}MCS: 17B10, 81R10}

\section{Introduction}

  Given a module of a Lie algebra~$\mf{g}$, it is an important and natural
  question to ask how this module decomposes under restriction of the
  action to a subalgebra~$\mf{p}$. This decomposition is
  described by non-negative integer numbers, the so-called branching
  coefficients. The aim of this paper is to provide new tools for
  determining branching coefficients in the case where both~$\mf{p}$
  and~$\mf{g}$ are finite-dimensional semi-simple Lie algebras.
  Several techniques have been developed to deal with this question.
  Among them are the use of generating functions, Schur functions
  and a generalization of Kostant's multiplicity formula as well as
  different kinds of algorithms. For details we refer the reader
  to~\cite{McKayPateraRand, GoodmanWallach:1990, FuchsSchweigert,
  FrancescoCFT,Klimyk:1967} and references therein.

  In this paper we develop a new approach which uses the fact that
  a semi-simple Lie algebra~$\mf{g}$ is naturally embedded in its affine
  extension~$\mf{\hat{g}}$. This makes available the powerful techniques
  of affine Kac-Moody algebras (see e.g.~\cite{Kac:1990}) and conformal field
  theories related to such algebras (see~\cite{FrancescoCFT} for instance).
  To give an example, we remind the reader that Verlinde's
  formula~\cite{Verlinde:1988sn} for
  fusion coefficients in $\mf{\hat{g}}_k$ Wess-Zumino-Witten (WZW) theories
  gives a generalization
  of the concept of tensor product coefficients of~$\mf{g}$.
  We will show that analogous relations hold for branching coefficients
  if we extend either~$\mf{g}$ or its subalgebra~$\mf{p}$ to the
  corresponding affine Kac-Moody algebra. In particular, in the first case
  there exists a relation to the theory of conformal boundary conditions
  and to the theory of fusion rings in WZW models~\cite{Quella:Other}.

  The paper is organized as follows. In Section~\ref{sc:Formula} we first
  provide some background on semi-simple Lie algebras
  and their affine extensions. Subsequently, we present a closed formula for
  branching coefficients based on the extension of the subalgebra~$\mf{p}$
  to~$\mf{\hat{p}}_k$. This formula is used in turn to give a simple derivation
  of a Racah-Speiser like algorithm in Section~\ref{sc:Algorithm}. Our results
  are applied to derive properties of branching coefficients and specialized
  to tensor product coefficients in Section~\ref{sc:Applications}.
  In addition we present a general procedure to obtain integral
  representations for branching coefficients. As an illustration of this
  method,
  we derive an integral representation for branching coefficients of the
  diagonal embedding~$A_1\hookrightarrow A_1\oplus A_1$.
  In Section~\ref{sc:NIMreps}, we consider a different approach based on
  representations of the fusion ring in $\mf{\hat{g}}_k$ WZW models.
  This leads to a Verlinde-like formula for branching coefficients
  and induces a second type of integral representations. We exploit the latter
  to obtain an explicit nontrivial integral representation for the branching
  coefficients of~$A_1\hookrightarrow A_2$ with embedding index~$1$.
  In addition we indicate that for the $A_{2n}$ series the fusion ring
  representation contains informations about two different
  embeddings at the same time. This solves some puzzle which remained open
  in~\cite{Quella:Other}.

\section{\label{sc:Formula}A closed formula for branching coefficients}

  We want to describe an embedding~$\mf{p}\hookrightarrow\mf{g}$ of one
  finite-dimensional semi-simple Lie algebra into another. For notational
  simplicity let us assume that~$\mf{p}$ actually is a simple Lie algebra
  but this does not restrict the validity of our results.
  Denote the weight
  lattices of~$\mf{p}$ and~$\mf{g}$ by~$\bar{L}_w$ and~$L_w$, respectively.
  Here and in what follows we will always use the convention
  that~$i,j,\cdots\in L_w$ and~$a,b,\cdots\in\bar{L}_w$.
  The finite-dimensional irreducible representations of the Lie algebras
  $\mf{p}$ and $\mf{g}$ are in one-to-one correspondence
  to the weights with non-negative integral Dynkin labels. These sets of
  so-called integrable highest weights of~$\mf{p}$ are denoted
  by~$\bar{P}^+\cong\bar{L}_w/W_{\mf{p}}$ with Weyl group~$W_{\mf{p}}$
  and similarly for~$\mf{g}$. Let~$M_a$ and~$M_i$ be the weight systems of
  the representations~$a\in\bar{P}^+$ and~$i\in P^+$ including the
  multiplicities.
  The embedding can be characterized by a projection
  $\mc{P}:\langle L_w\rangle\to\langle\bar{L}_w\rangle$ where
  $\langle L\rangle$ means the span of the lattice~$L$
  over~$\Complex$. Under this projection, the weight system~$M_i$ of the
  representation~$i\in P^+$ of~$\mf{g}$ decomposes into weight systems of
  representations of~$\mf{p}$ according to
\begin{equation}
  \label{eq:ModuleDecomposition}
  \mc{P}M_i=\bigoplus_{a\in\bar{P}^+} {b_i}^a M_a\quad.
\end{equation}
  The numbers~${b_i}^a\in\Natural_0$ are called branching coefficients.
  Our aim is to find an explicit and general formula for the
  coefficients~${b_i}^a$
  with~$i\in P^+$ and~$a\in\bar{P}^+$. To achieve this, we consider the
  untwisted affine extension~$\hat{\mf{p}}_k$ of~$\mf{p}$. The level~$k$
  has to be chosen large enough and depends on the value of~$i$.
  This statement will be made precise below.
  The integrable highest weights of $\hat{\mf{p}}_k$
  are given by the set
  $\bar{P}_k^+=\bar{L}_w/(W_{\mf{p}}\ltimes k\bar{L}^\vee)$ where we used the
  decomposition of the affine Weyl group into a semi-direct product of finite
  Weyl group and translations by~$k$ times the coroot lattice~$\bar{L}^\vee$.
  If we introduce the
  notation~$k(c)=(\theta,c)_{\mf{p}}$ where~$\theta$ is the highest root
  of~$\mf{p}$ we may write $\bar{P}_k^+=\{a\in\bar{P}^+|k(a)\leq k\}$.
  The bracket $(\cdot,\cdot)_{\mf{p}}$
  denotes the scalar product on the weight space~$\langle\bar{L}_w\rangle$
  which is induced by the Killing form.
  It is given in terms of the quadratic form matrix~$F_{\mf{p}}$ if
  the weights are written using Dynkin labels, i.e.
  $(\lambda,\mu)_{\mf{p}}=\lambda^TF_{\mf{p}}\mu$.
  In the following we will always identify in a natural way an integrable
  highest weight representation $\hat{c}\in P_k^+$
  of~$\hat{\mf{p}}_k$ with a highest weight~$c\in\bar{P}^+$
  of~$\mf{p}\hookrightarrow\hat{\mf{p}}_k$.

  Before we continue, let us briefly
  introduce further objects that will be needed as we proceed.
  The character of an highest weight representation~$i\in P^+$
  of $\mf{g}$ is defined as
\begin{equation}
  \label{eq:CharacterDef}
  \chi_i(\cdot)=\sum_{j\in M_i}\:e^{(j,\cdot)_{\mf{g}}}
\end{equation}
  and analogously for~$\mf{p}$.
  The second ingredient of our formula
  is the modular S~matrix of~$\hat{\mf{p}}_k$ which, for $a,b\in\bar{P}_k^+$,
  is given by the Kac-Peterson formula~\cite{Kac:1990}
\begin{equation}
  \label{eq:SMatrixDef}
  S_{ab}
  =i^{|\Delta_+|}|\:\bar{L}_w/\bar{L}^\vee|^{-1/2}(k+g^\vee)^{-r/2}
   \sum_{w\in W}\epsilon(w)\exp\Bigl\{-\frac{2\pi i}{k+g^\vee}\Bigl(w(a+\rho),b+\rho\Bigr)_{\mf{p}}\Bigr\}\quad.
\end{equation}
  This formula involves the rank of the Lie algebra~$r$, the number of
  positive roots~$|\Delta_+|$, the Weyl vector~$\rho$, the dual
  Coxeter number $g^\vee=(\theta,\rho)_{\mf{p}}+1$ and a sum over the
  Weyl group~$W$ including its sign function $\epsilon$.
  We omit the index~$\mf{p}$ because we will not encounter the
  corresponding objects for the Lie algebra~$\mf{g}$.
  Due to Weyl's character formula we may write
\begin{equation}
  \label{eq:SMatrixCharacterRelation}
  \chi_a(\xi_b)=\frac{S_{ba}}{S_{b0}}
  \hspace{1cm}\text{ where }\hspace{1cm}
  \xi_b=-\frac{2\pi i}{k+g^\vee}(b+\rho)
  \hspace{0.5cm}\text{ and }\hspace{0.5cm}
  a,b\in\bar{P}_k^+\quad.
\end{equation}

  We are now prepared to state the first result of this paper.
\begin{theorem}
  \label{thm:BranchingFormula}
  Consider an embedding~$\mf{p}\hookrightarrow\mf{g}$ of two
  finite-dimensional semi-simple Lie algebras.
  Let~$\mc{P}:\langle L_w\rangle\to\langle\bar{L}_w\rangle$
  be the projection matrix characterizing the embedding
  and~$a\in\bar{P}^+,i\in P^+$ be two arbitrary but integrable highest
  weights. Define a
  map~$\mc{P}^\ast=F_{\mf{g}}^{-1}\mc{P}^TF_{\mf{p}}:
  \langle\bar{L}_w\rangle\to\langle L_w\rangle$
  and let~$k$ be a number such that $k\geq\max\{k(c)|{b_i}^c\neq0\}$.
  Then we have
\begin{equation}
  \label{eq:BranchingFormula}
  {b_i}^a
  =\sum_{d\in\bar{P}_k^+}\sum_{j\in M_i}\bar{S}_{da}S_{d0}e^{-\frac{2\pi i}{k+g^\vee}(\mc{P}j,d+\rho)_{\mf{p}}}
  =\sum_{d\in\bar{P}_k^+}\bar{S}_{da}S_{d0}\chi_i(\mc{P}^\ast\xi_d)\quad.
\end{equation}
\begin{proof}
  For notational simplicity we assume~$\mf{p}$ to be simple.
  Let us first note that $\max\{k(c)|{b_i}^c\neq0\}$ exists as all weight
  systems involved are finite.
  We then start by writing down the identity
\begin{equation}
  \label{eq:StartingPoint}
  \sum_{c\in\bar{P}_k^+}{b_i}^c\:\frac{S_{dc}}{S_{d0}}
  =\sum_{c\in\bar{P}^+}{b_i}^c\:\chi_c(\xi_d)=\chi_i(\mc{P}^\ast\xi_d)\quad.
\end{equation}
  If we multiply both sides of~\eqref{eq:StartingPoint} with
  $\bar{S}_{da}S_{d0}$ and sum over all $d\in\bar{P}_k^+$ we obtain
  the desired result due to the
  unitarity~$\sum_{d}\bar{S}_{da}S_{dc}=\delta_c^a$ of the S~matrix.
  Thus we only have to motivate~\eqref{eq:StartingPoint}.
  The left equality simply results from~\eqref{eq:SMatrixCharacterRelation}
  and the condition on the level~$k$,
  but the right equality is more interesting.
  Let~$M_i$ be the weight system of the representation~$i$ including
  all multiplicities. We insert the definition~\eqref{eq:CharacterDef} of the
  characters into~\eqref{eq:StartingPoint}.
  After this substitution, the sum
  on the right hand side of~\eqref{eq:StartingPoint} is over~$M_i$ and
  involves scalar products~$(j,\cdot)_{\mf{g}}$. In contrast to this,
  the sum in the middle is over the projected weights $\mc{P}M_i$
  and therefore involves scalar products of the
  form~$(\mc{P}j,\cdot)_{\mf{p}}$. The sum in both cases runs essentially
  over the same set~$M_i$. Therefore the equality in~\eqref{eq:StartingPoint}
  holds if we can identify the scalar products according to
  $(\mc{P}j,\cdot)_{\mf{p}}=(j,\mc{P}^\ast\cdot)_{\mf{g}}$.
  Writing this relation in terms of quadratic form matrices, we see
  that~$\mc{P}^\ast$ was constructed exactly in a way that this identity holds.
\end{proof}
\end{theorem}

  Notice the following remarkable observation. If we
  could rewrite $\mc{P}^\ast\xi_d$ as~$\xi_j^\prime$ for some integrable
  highest weight~$j$ of~$\hat{\mf{g}}_{k^\prime}$ at a certain
  level~$k^\prime$, we could apply eq.~\eqref{eq:SMatrixCharacterRelation}
  and eq.~\eqref{eq:BranchingFormula}
  would reduce to a Verlinde-like formula~\cite{Verlinde:1988sn}
  for branching coefficients. In general, this does not seem to be
  possible because $F_{\mf{g}}^{-1}$ might cause negative entries
  in~$\mc{P}^\ast$. We will see however in Section~\ref{sc:NIMreps}
  that in some specific cases we are able to recover a Verlinde-like
  formula using a different approach. 
  \medskip
  
  Let us briefly comment on the changes if~$\mf{p}$ is finite-dimensional
  and semi-simple but not simple. Under these circumstances we have a
  decomposition $\mf{p}\cong\oplus_{s=1}^n \mf{p}_s$ of~$\mf{p}$
  into simple Lie algebras~$\mf{p}_s$. In the affine extension, each
  simple factor obtains its own level:
  $\hat{\mf{p}}_k\cong\oplus_{s=1}^n (\hat{\mf{p}}_s)_{k_s}$ with
  $k=(k_1,\ldots,k_n)$. All relevant structures like
  the weight lattice, the Weyl group, the quadratic form matrix and
  the modular S~matrix 'factorize' in some sense, i.e. they are given
  by a direct sum, a product, a block diagonal matrix or factorize
  in the original sense of the word. Obviously, the proof of
  theorem~\ref{thm:BranchingFormula} still remains valid if one takes
  these notational difficulties into account. In particular,
  the condition $k\geq\max\{k(c)|{b_i}^c\neq0\}$ actually means
  $k_s\geq\max\{k_s(c)|{b_i}^c\neq0\}$ in this case.

\section{\label{sc:Algorithm}An alternative derivation of a Racah-Speiser
         like algorithm for branching rules}

  We will now use formula~\eqref{eq:BranchingFormula} to give an easy
  derivation of a well-known algorithm~\cite{Klimyk:1967} for the
  calculation of branching
  coefficients which is the basis of many computer algebra
  programs\footnote{I am grateful to M. van Leeuwen for providing this
  information.}.
  The algorithm exhibits some similarity with the Racah-Speiser
  algorithm for the calculation of tensor product multiplicities
  (see also~\cite{Kac:1990,Walton:1990sc,Walton:1990qs,Fuchs:1990hm,
  Furlan:1990ce,GoodmanWenzl}
  for its extension to fusion rules).
\begin{theorem}
  \label{thm:RacahSpeiser}
  Consider an embedding~$\mf{p}\hookrightarrow\mf{g}$ of finite-dimensional
  semi-simple Lie algebras. Let~$i\in P^+$ be a highest weight of~$\mf{g}$ and
  $\mc{P}:\langle L_w\rangle\to\langle\bar{L}_w\rangle$
  be the projection matrix characterizing
  the embedding. The decomposition $\mc{P}M_i=\oplus_{a}{b_i}^aM_a$ can be
  obtained by the following algorithm\footnote{The algorithm and the proof
  are based on~\cite{Schweigert:Unpublished} in which a slightly
  different algorithm for calculating NIM-reps for twisted boundary conditions
  in WZW models is proved.}.
\begin{enumerate}
\item Calculate the weight system of the representation~$i$ including the
  multiplicities. This gives some set~$M_i\subset L_w$.
\item Project this set to~$\bar{L}_w$ and add the Weyl vector
  of the subalgebra~$\mf{p}$.
  Now we are dealing with the set $Z_i=\mc{P}M_i+\rho\subset\bar{L}_w$
  including the multiplicities.
\item For each weight of~$Z_i$ use a Weyl reflection to map it into
  the fundamental Weyl chamber where all Dynkin labels are non-negative.
  An algorithm in terms of elementary Weyl reflections can be found
  in~\cite{FuchsSchweigert} for example.
\item Drop all weights lying on the boundary of the fundamental Weyl
  chamber and subtract the Weyl vector~$\rho$ of the subalgebra~$\mf{p}$
  from the remaining ones.
\item Add up all these contributions including the signs of the
  relevant Weyl reflections and the multiplicities. The coefficient obtained
  for each weight~$a\in\bar{P}^+$ is just the number~${b_i}^a$.
\end{enumerate}
\begin{proof}
  Again we assume~$\mf{p}$ to be simple without loss of generality.
  Essentially, the idea is to evaluate
  equation~\eqref{eq:BranchingFormula} for~$k\to\infty$.
  We insert the definitions~\eqref{eq:CharacterDef},\eqref{eq:SMatrixDef}
  for the characters and the S~matrix.
  Denoting the prefactor by
  $\mc{N}=|\bar{L}_w/\bar{L}^\vee|^{-1}(k+g^\vee)^{-r}$ we obtain
\begin{equation}
  \label{eq:ProofOne}
  {b_i}^a
  =\mc{N}\sum_{d\in \bar{P}_k^+}\sum_{w_1,w_2\in W}\sum_{j\in M_i}
   \epsilon(w_1)\epsilon(w_2)
   \exp\Bigl\{-\frac{2\pi i}{k+g^\vee}\Bigl(\mc{P}j+w_1\rho-w_2(a+\rho)\:,\:d+\rho\Bigr)_{\mf{p}}\Bigr\}
\end{equation}
  where we already made use of the defining relation
  $(j,\mc{P}^\ast\xi_d)_{\mf{g}}=(\mc{P}j,\xi_d)_{\mf{p}}$ for~$\mc{P}^\ast$.
  The next step consists in evaluating the sum over~$d$. We define a function
  $f(d)$ by ${b_i}^a=\sum_{d\in \bar{P}_k^+}f(d+\rho)$. The function $f(c)$
  as read of from eq.~\eqref{eq:ProofOne} has two important properties. First,
  it satisfies $f(wc)=f(c)$ for all $w\in W$. Indeed, the Weyl reflection
  may be absorbed into a redefinition\footnote{Note that the weight system
  which belongs to an arbitrary representation is invariant under Weyl
  transformations. In particular this holds for the set~$\mc{P}M_i$.}
  of~$w_1,w_2$ and~$j$. To derive the second property let us define the set
  $\bar{P}_{k+g^\vee}^{++}=\bar{P}_k^++\rho$. It turns out that
  $\bar{P}_{k+g^\vee}^{++}$ exactly contains the elements of
  $\bar{P}_{k+g^\vee}^{+}$ which do not lie at the boundary of the
  corresponding affine Weyl chamber. This boundary is given by the
  set of all weights which are invariant under at least one elementary Weyl
  reflection including the shifted reflection at the $k$-dependent hyperplane
  described by $(\theta,\cdot)_{\mf{p}}=k+g^\vee$. One may show that
  $f(c)=0$ if $c$ is invariant under an affine fundamental Weyl reflection.
  To see this, note that the function $g_x(c)=S_{x,c-\rho}$ which enters
  $f(d)$ satisfies $g_x(\hat{w}c)=\epsilon(\hat{w})g_x(c)$ with respect to
  any affine Weyl transformation $\hat{w}\in W\ltimes(k+g^\vee)L^\vee$.
  These considerations lead to the simple relation
\begin{equation}
  \label{eq:ProofTwo}
  {b_i}^a
  =\frac{1}{|W|}\sum_{d\in\bar{P}_k^+}\sum_{w\in W}f\bigl(w(d+\rho)\bigr)
  =\frac{1}{|W|}\sum_{c\in\bar{P}_{k+g^\vee}^+}\sum_{w\in W}f\bigl(wc\bigr)
  =\frac{1}{|W|}\sum_{c\in L_w/(k+g^\vee)L^\vee}f(c)\quad.
\end{equation}
  We are now in a situation where we are able to perform the sum over
  $c\in L_w/(k+g^\vee)L^\vee$. The sum over the exponentials in
  eq.~\eqref{eq:ProofOne} exactly gives a non-vanishing result if
  $\mc{P}j+w_1\rho-w_2(a+\rho)\in(k+g^\vee)L^\vee$. In this case
  it obviously compensates the normalization factor $\mc{N}$. In the
  limit $k\to\infty$ this condition reduces to a Kronecker symbol and
  we are left with the $k$-independent expression
\begin{equation}
  \label{eq:ProofFour}
  {b_i}^a
  =\frac{1}{|W|}\sum_{w_1\in W}\sum_{w_2\in W}\sum_{j\in M_i}
   \epsilon(w_1)\epsilon(w_2)
   \delta_{w_2(a+\rho),\mc{P}j+w_1\rho}\quad.
\end{equation}
  Next shift~$w_2$ to the other side of the Kronecker symbol
  ($w_2^{-1}=w_2$) and
  resum~$w_1\mapsto w_2w_1$ as well as~$\mc{P}j\mapsto w_2w_1\mc{P}j$. The
  expression under the sum then obviously does not depend on~$w_2$ anymore.
  By summing over~$w_2$, we compensate the factor~$1/|W|$.
  The final result is
\begin{equation}
  \label{eq:ProofFive}
  {b_i}^a
  =\sum_{j\in M_i}\sum_{w\in W}
   \epsilon(w)\delta_{a,w(\mc{P}j+\rho)-\rho}\quad.
\end{equation}
  For each weight~$\mc{P}j+\rho$ lying at the boundary
  of a Weyl chamber there always exists an elementary Weyl reflection which
  leaves it fixed. These weights may be omitted because they would
  contribute twice with different sign. Inserting our result
  into equation~\eqref{eq:ModuleDecomposition} proves the theorem.
\end{proof}
\end{theorem}

\section{\label{sc:Applications}Applications and an integral formula for
         branching coefficients}

  Using theorem~\ref{thm:BranchingFormula} and
  formula~\eqref{eq:BranchingFormula} one may explicitly check
  some well known properties of branching coefficients. Thus one obtains
\begin{corollary}
  \label{cor:Properties}
  Let $\mf{h}\hookrightarrow\mf{p}\hookrightarrow\mf{g}$ be
  an embedding of finite-dimensional semi-simple Lie algebras and
  denote the integrable highest weights by $\alpha,\beta,\ldots$
  and $a,b,\ldots$ and $i,j,\ldots$ respectively.
  The branching coefficients have the following properties.
\begin{enumerate}
\item The trivial representation~$0\in P^+$ decomposes according
  to ${b_0}^a=\delta_0^a$.
\item Denoting the conjugate representation by~$(\cdot)^+$, the
  relation~${b_{i^+}}^{a^+}={b_i}^a$ holds.
\item The branching coefficients of the embedding
  $\mf{h}\hookrightarrow\mf{p}\hookrightarrow\mf{g}$ are related by
  ${b_i}^\alpha=\sum_a {b_i}^a{b_a}^\alpha$.
\item In the decomposition of a tensor product $V_i\otimes V_j$ both reductions
  are equivalent, i.e. the branching coefficients satisfy
  $\sum_l{N_{ij}}^l{b_l}^a=\sum_{c,d}{b_i}^c{b_j}^d{N_{cd}}^a$.
\end{enumerate}
\begin{proof}
  The first relation holds because~$\chi_0(\cdot)=1$.
  For the second relation one
  needs that the charge conjugation matrix satisfies~$C=C^T=C^{-1}$ as well as
  $F\circ C=C\circ F$ and $C_{\mf{p}}\circ\mc{P}=\mc{P}\circ C_{\mf{g}}$.
  The third relation is due to the fact
  that~$\mc{P}^\ast(\mf{h}\hookrightarrow\mf{p}\hookrightarrow\mf{g})
  =\mc{P}^\ast(\mf{p}\hookrightarrow\mf{g})\circ
  \mc{P}^\ast(\mf{h}\hookrightarrow\mf{p})$.
  The last property can be checked using the Verlinde formula
  for~${N_{cd}}^a$ (this is valid if we choose~$k$ large enough, see
  corollary~\ref{cor:Tensor}), the unitarity of the S~matrix
  and the property $\chi_i\chi_j=\sum_{l}{N_{ij}}^l\chi_l$ of characters.
\end{proof}
\end{corollary}

  The diagonal embedding $\mf{g}\hookrightarrow\mf{g}\oplus\mf{g}$ is special
  in the sense that its branching coefficients correspond to the tensor
  products in~$\mf{g}$. In this case theorem~\ref{thm:BranchingFormula} implies
\begin{corollary}
  \label{cor:Tensor}
  Let~$\mf{g}$ be a finite dimensional semi-simple Lie algebra and
  $V_i,V_k$ two fixed integrable highest weight modules. There exists
  some $k_0\in\Natural$ such that  the coefficients
  in the decomposition $V_i\otimes V_j=\oplus_l {N_{ij}}^l V_l$ may be
  expressed by the Verlinde formula
\begin{equation*}
  {N_{ij}}^l=\sum_{m\in P_k^+}\frac{\bar{S}_{ml}S_{mi}S_{mj}}{S_{m0}}
\end{equation*}
  for all integers $k>k_0$.
\begin{proof}
  This is a simple consequence of theorem~\ref{thm:BranchingFormula}
  and the fact that the branching coefficients
  for the diagonal embedding $\mf{g}\hookrightarrow\mf{g}\oplus\mf{g}$
  with projection $\mc{P}(l_1,l_2)=l_1+l_2$
  are given by the tensor product multiplicities of~$\mf{g}$. Using the
  definition one obtains $\mc{P}^\ast(l)=(l,l)$.
  The character of $\mf{g}\oplus\mf{g}$ in~\eqref{eq:BranchingFormula}
  decomposes into a product
  of two characters of~$\mf{g}$ with argument~$\xi_l$. Applying
  equation~\eqref{eq:SMatrixCharacterRelation} gives the desired result.
\end{proof}
\end{corollary}

  The last remarks concern integral formulae for branching coefficients
  which may be deduced from theorem~\ref{thm:BranchingFormula}. We will
  not give a proof that this is always possible but only give the idea
  and a simple example for illustration. First we observe
  that the S~matrices and the character
  in~\eqref{eq:BranchingFormula} both have a dependence
  $\sim(d+\rho)/(k+g^\vee)$ on the summation index~$d$. In addition, the two
  S~matrices give a total prefactor of the form~$(k+g^\vee)^{-r}$
  where~$r$ is the rank of the subalgebra, i.e. the number of
  independent components of~$d$. Therefore it is likely
  that in many (if not all) cases we may rewrite the sum as an integral
  in the limit $k\to\infty$ and in this way recover an integral
  representation of branching coefficients.
  \medskip
  
  We show how this works in a very simple example and rederive some integral
  formula for the (of course well-known) tensor product multiplicities of
  representations of~$A_1$, i.e. the branching rules of the diagonal
  embedding~$A_1\hookrightarrow A_1\oplus A_1$.
  The characters of~$A_1$ read
  $\chi_a(x)=\sinh\frac{x}{2}(a+1)/\sinh\frac{x}{2}$ and
  the S~matrix is given by
  $S_{ab}=\sqrt{\frac{2}{k+2}}\sin\frac{\pi}{k+2}(a+1)(b+1)$.
  Using the factorization of the $A_1\oplus A_1$-character,
  equation~\eqref{eq:BranchingFormula} implies for all~$k$ greater than some~$k_0$
\begin{eqnarray*}
  {N_{a_1a_2}}^{a}&=&{b_{(a_1,a_2)}}^a\\
  &=&\frac{2}{k+2}\sum_{b=0}^k\frac{\sin\frac{\pi}{k+2}(a+1)(b+1)\:\sin\frac{\pi}{k+2}(a_1+1)(b+1)\:\sin\frac{\pi}{k+2}(a_2+1)(b+1)}{\sin\frac{\pi}{k+2}(b+1)}\\
  &=&2\int_0^1\hspace{-0.6em}dx\quad\frac{\sin\pi(a+1)x\:\sin\pi(a_1+1)x\:\sin\pi(a_2+1)x}{\sin\pi x}\quad.
\end{eqnarray*}
  For the last equality we consider the sum to be a Riemann sum with an
  equidistant partition of the interval
  $[1/(k+2),(k+1)/(k+2)]$ into intervals of length $\Delta x=1/(k+2)$.
  Due to continuity we may
  extend the interval to~$[0,1]$. As the integral exists, it is given by the
  previous series in the limit~$k\to\infty$.
  While such integral representations for general branching coefficients
  seem to be new, similar statements for tensor products can
  for example be found in~\cite[p.~534]{FrancescoCFT}.

\section{\label{sc:NIMreps}Relation to conformal field theory 
  and a Verlinde-like formula for branching coefficients}

  Let us mention that there exists an interesting relation of our work
  to the classification of boundary conditions in  a special class
  of conformal field theories~\cite{FrancescoCFT}, the so-called WZW
  models with affine symmetry~$\hat{\mf{g}}_k$.
  It can be shown that to every consistent set of conformal boundary
  conditions there exists a so-called NIM-rep of the corresponding
  fusion ring~\cite{Behrend:1999bn}.
  A NIM-rep is given by non-negative integral
  matrices~${\bigl(n_i^{(k)}\bigr)_b}^a$
  satisfying $n_i^{(k)}n_j^{(k)}=\sum_l N_{ij}^{(k)l} n_l^{(k)}$
  and~$n_{i^+}^{(k)}=\bigl(n_i^{(k)}\bigr)^T$ where
  the numbers~$N_{ij}^{(k)l}$ are the fusion rules of the model.
  One can show that every NIM-rep (at least the finite ones)
  can be diagonalized by an unitary matrix~$U$ and one obtains a
  Verlinde-like formula of the form
\begin{equation}
  \label{eq:VerlindeFormula}
  {\Bigl(n_i^{(k)}\Bigr)_b}^a
  =\sum_d\frac{\bar{U}_{ad}U_{bd}S_{i\phi(d)}}{S_{0\phi(d)}}
\end{equation}
  with some map $\phi:\{a,b,c,d,\ldots\}\to P_k^+$.
  For recent work on NIM-reps and the connection to the
  classification of conformal boundary conditions
  see~\cite{Behrend:1999bn,Gannon:2001ki}.
  Explicit formulae for~$U$ may be found in~\cite{Fuchs:1999zi,
  Birke:1999ik}. An approach based on graphs is given in~\cite{Behrend:1999bn}.
  Note that not all NIM-reps have physical significance~\cite{Gannon:2001ki}.
  \smallskip
  
  We now want to show how our construction is related to the theory of
  NIM-reps. Let~$\mf{p}$ be a subalgebra of~$\mf{g}$.
  Denote the tensor product multiplicities of~$\mf{p}$ by~${N_{ab}}^c$ and the
  branching coefficients by~${b_i}^a$.
  One can easily show that the
  matrices~${\bigl(n_i\bigr)_b}^a=\sum_c {b_i}^c {N_{cb}}^a$
  constitute a NIM-rep of the fusion ring of the WZW model 
  associated with~$\mf{\hat{g}}_k$ at level~$k\to\infty$.
  In this limit the fusion rules~${N_{ij}}^l=\lim_{k\to\infty}N_{ij}^{(k)l}$
  reduce to the tensor product multiplicities of~$\mf{g}$.
  The proof of the NIM-rep properties relies on the fact that the two
  possibilities of decomposing a module~$V_i\otimes V_j$ of~$\mf{g}$ into
  modules of~$\mf{p}$ are equivalent
  (compare corollary~\ref{cor:Properties}) and on the associativity of tensor
  products. It is easy to generalize the considerations of the
  Sections~\ref{sc:Formula} and~\ref{sc:Algorithm} to obtain
\begin{equation}
  \label{eq:NIMRep}
  {\bigl(n_i\bigr)_b}^a
  =\sum_{c\in\bar{P}^+} {b_i}^c {N_{cb}}^a
  =\sum_{d\in\bar{P}_k^+}\bar{S}_{da}S_{db}\chi_i(\mc{P}^\ast\xi_d)
  =\sum_{j\in M_i}\sum_{w\in W}
   \epsilon(w)\delta_{a,w(\mc{P}j+b+\rho)-\rho}
\end{equation}
  for sufficiently large values of the level~$k$.
  Note that we did not rely on methods of conformal field theory to obtain
  this result. We just provided a completely algebraic treatment along the
  lines of the first four Sections.
  \smallskip

  Our next task is to relate the purely algebraic NIM-reps of the last
  paragraph to results from conformal field theory. Indeed, one may
  prove~\cite{Quella:Other} that NIM-reps which come along with certain kinds
  of boundary conditions\footnote{These so-called twisted boundary conditions
  are connected to non-trivial symmetries of the Dynkin diagram of~$\mf{g}$.}
  in~$\mf{\hat{g}}_k$ WZW theories coincide with the
  expressions given in~\eqref{eq:NIMRep} in the limit~$k\to\infty$. This means
  that NIM-reps~${\bigl(n_i^{(k)}\bigr)_b}^a$ which may be described as in
  equation~\eqref{eq:VerlindeFormula} for finite values of~$k$, reduce to the
  expression~\eqref{eq:NIMRep} in the limit~$k\to\infty$ for certain
  distinguished subalgebras~$\mf{p}$. In particular, this holds true for the
  special matrix elements~${b_i}^a={\bigl(n_i\bigr)_0}^a$. 
  Starting from~\eqref{eq:VerlindeFormula},
  we thus obtain another representation of branching coefficients for these
  distinguished embeddings. On one hand this yields another version of a
  Racah-Speiser like algorithm~\cite{Schweigert:Unpublished} invented
  originally for the calculation of NIM-reps. On the other hand it may be
  used to derive alternative integral representations for branching
  coefficients along the lines of Section~\ref{sc:Applications} if one
  takes the explicit expressions for the matrices~$U$ (see for
  example~\cite{Birke:1999ik}) and the results of~\cite{Quella:Other}
  into account.
  \smallskip

\begin{table}
  \centerline{\begin{tabular}{c|ccccccc}
  $\mf{g}$ & $A_2$ & $A_{2n-1}$ & $A_{2n}$ & $A_{2n}$ & $D_4$ & $D_n$ & $E_6$ \\\hline
  $\mf{p}$ & $A_1\:(x_e=1,4)$ & $C_n$ & $C_n\hookrightarrow A_{2n-1}$ & $(B_n)$ & $G_2\hookrightarrow B_3$ & $B_{n-1}$ & $F_4$
  \end{tabular}}
  \caption{\label{tb:Embeddings}Embeddings of simple Lie algebras, known to
  be related to the limit $k\to\infty$ of NIM-reps of WZW models. The relevant
  subalgebra is specified by a sequence of maximal embeddings. The
  statement for the embedding $B_n\hookrightarrow A_{2n}$ is based on a
  conjecture only and not yet established rigorously.}
\end{table}

  Table~\ref{tb:Embeddings} contains a list of embeddings to which these
  considerations are known or conjectured to be applicable. A large part of
  these identifications are taken from~\cite{Quella:Other}. Note, that the
  corresponding subalgebra in almost all examples is given by the subalgebra
  invariant under the Lie algebra automorphism induced by the Dynkin diagram
  symmetry to which the NIM-rep belongs. It remained obscure, however, why in
  the case of $\mf{g}=A_{2n}$
  the relevant subalgebra is given by $C_{n}$  (the so-called orbit Lie
  algebra~\cite{Fuchs:1996zr} of~$A_{2n}$) and not by the subalgebra~$B_n$,
  invariant with respect to the non-trivial diagram automorphism of $A_{2n}$.
  Below we will partly fill this gap and show that one and
  the same NIM-rep may lead to two {\em different} subalgebras under two
  distinct identifications of NIM-rep labels. We will prove this remarkable
  feature of NIM-reps in the case of $A_2$ and comment on the case of
  $A_{2n}$ with $n>1$ afterwards. It is an open problem whether all NIM-reps of
  the type~${\bigl(n_i\bigr)_b}^a=\sum_c {b_i}^c {N_{cb}}^a$ may be extended
  to finite
  values of~$k$. This is certainly true for NIM-reps related to the
  embeddings given in Table~\ref{tb:Embeddings} (with some caveat regarding
  embeddings of the type $B_n\hookrightarrow A_{2n}$ for $n>1$) or to diagonal
  embeddings $\mf{g}\hookrightarrow\mf{g}\oplus\mf{g}$, but to our
  knowledge nothing is known for arbitrary
  embeddings~$\mf{p}\hookrightarrow\mf{g}$.
  \medskip

  Let us illustrate our considerations with an example. The Lie
  algebra~$\mf{g}=A_2$ has exactly one automorphism $\omega$ related to a
  non-trivial Dynkin diagram symmetry, where it acts as a permutation of nodes.
  On the level of weights it thus acts as a permutation of Dynkin labels
  $\omega(a_1,a_2)=(a_2,a_1)$. As is well known, $\omega$ induces a conformal
  boundary condition in the $\bigl(A_2^{(1)}\bigr)_k$ WZW model.
  Following~\cite{Birke:1999ik} the boundary labels are given by half-integer
  symmetric weights $\alpha,\beta=(0,0),(1/2,1/2),\cdots,
  (\lfloor k/2\rfloor/2,\lfloor k/2\rfloor/2)$. Here, the symbol
  $\lfloor x\rfloor$ denotes the largest integer number smaller or equal
  to~$x$. The relevant NIM-reps
\begin{equation}
  \label{eq:NIMrepA1}
  {\Bigl(n_{(i_1,i_2)}^{(k)}\Bigr)_\beta}^\alpha
  =\sum_{\mu=0}^{\lfloor k/2\rfloor}\frac{\bar{S}_{\mu\alpha}^\omega S_{\mu\beta}^\omega S_{(\mu,\mu),(i_1,i_2)}}{S_{(\mu,\mu),(0,0)}}
\end{equation}
  may be calculated~\cite{Birke:1999ik} using the explicit formula
\begin{equation}
  \label{eq:TSMatrixA2}
  S_{\mu\alpha}^\omega
  =\frac{2}{\sqrt{k+3}}\sin\frac{2\pi}{k+3}(\mu+1)(2\alpha+1)
\end{equation}
  where we identified the tupel~$\alpha$ with one of its (identical) entries.
  The obvious similarity of this expression with the S matrix of $A_1^{(1)}$
  in mind we may ask whether the NIM-rep~\eqref{eq:NIMrepA1} in the limit
  $k\to\infty$ reduces to a NIM-rep of the type~\eqref{eq:NIMRep} coming from
  an embedding $A_1\hookrightarrow A_2$. To check this assertion we have to
  identify the half-integer symmetric NIM-rep label $\alpha,\beta$ with
  weights $a,b$ of $A_1$ via some map
  $\Psi:\{a,b,\ldots\}\to\{\alpha,\beta,\ldots\}$. Unfortunately there are two
  of these embeddings at our disposal and we have to worry
  which is the correct one. In~\cite{Quella:Other} a map~$\Psi$ has been
  proposed which leads to the embedding with
  projection~$\mc{P}(i_1,i_2)=i_1+i_2$ and
  embedding index $x_e=1$. We will show below, however, that there is another
  map~$\Psi^\prime$ yielding the embedding with
  projection~$\mc{P}^\prime(i_1,i_2)=2(i_1+i_2)$ and embedding index
  $x_e^\prime=4$.
  \smallskip

  We will discuss the first case first and derive an
  integral representation for branching coefficients of the embedding
  $A_1\hookrightarrow A_2$ with
  embedding index~$x_e=1$. In this case one has to use the identification map
  $\Psi(a)=(a/2,a/2)$~\cite{Quella:Other}. In order to be able to apply
  equation~\eqref{eq:NIMrepA1} we further need the special quotient
\begin{equation*}
  \frac{S_{(\mu,\mu),(i_1,i_2)}}{S_{(\mu,\mu),(0,0)}}
  =\frac{\sin\frac{2\pi}{k+3}(i_1+1)(\mu+1)+\sin\frac{2\pi}{k+3}(i_2+1)(\mu+1)-\sin\frac{2\pi}{k+3}(i_1+i_2+2)(\mu+1)}{8\sin^3\frac{\pi}{k+3}(\mu+1)\cos\frac{\pi}{k+3}(\mu+1)}
\end{equation*}
  of S~matrices of~$A_2$ which may be computed using the
  Kac-Peterson formula~\eqref{eq:SMatrixDef}. Following~\cite{Quella:Other}
  one may write
\begin{equation*}
  {b_{(i_1,i_2)}}^a
  =\lim_{k\to\infty}{{\Bigl(n_{(i_1,i_2)}^{(k)}\Bigr)}_{\Psi(0)}}^{\Psi(a)}
  =\lim_{k\to\infty}\sum_{\mu=0}^{\lfloor k/2\rfloor}\frac{\bar{S}_{\mu\Psi(a)}^\omega S_{\mu\Psi(0)}^\omega S_{(\mu,\mu),(i_1,i_2)}}{S_{(\mu,\mu),(0,0)}}\quad.
\end{equation*}
  Performing the continuum limit we arrive at
\begin{equation*}
  {b_{(i_1,i_2)}}^a
  =\int_0^{1/2}\hspace{-1em}dx\quad\frac{\sin2\pi(a+1)x\Bigl(\sin2\pi(i_1+1)x+\sin2\pi(i_2+1)x-\sin2\pi(i_1+i_2+2)x\Bigr)}{\sin^2\pi x}\quad.
\end{equation*}
  We thus obtained a non-trivial integral formula for the branching
  coefficients of the embedding $A_1\hookrightarrow A_2$ with embedding
  index $x_e=1$.
  \smallskip

  As stated above there is another identification of NIM-rep labels with
  weights of $A_1$, leading to the embedding $A_1\hookrightarrow A_2$
  with $x_e^\prime=4$. We will assume $k$ to be even in what follows. For any
  even weight $a$ of $A_1$ define $\Psi^\prime(a)=(k/4,k/4)-(a/4,a/4)$. Before
  we continue, let us mention
  two obvious differences compared to the previous identification
  map~$\Psi$. First, the identification map~$\Psi^\prime$ involves the
  level~$k$ explicitly. Second, the map is only well-defined for a subset of
  weights of $A_1$, i.e. the even ones. One may easily check however, that
  this restriction corresponds exactly to a general selection rule of the
  branching coefficients of $A_1\hookrightarrow A_2$ with $x_e^\prime=4$. We
  use our new identification map $\Psi^\prime$ to rewrite~\eqref{eq:TSMatrixA2}
  according to
\begin{equation*}
  S_{\mu a}^{\omega\prime}=S_{\mu\Psi^\prime(a)}^\omega
  =\frac{2(-1)^\mu}{\sqrt{k+3}}\sin\frac{\pi}{k+3}(\mu+1)(a+1)\quad.
\end{equation*}
  Apart from a factor $\sqrt{2}(-1)^\mu$ this is just the S matrix
  $S_{\mu a}^{A_1}$ of
  $A_1^{(1)}$ at level $k+1$. Using~\eqref{eq:SMatrixCharacterRelation} we
  are now able to write equation~\eqref{eq:NIMrepA1} as
\begin{equation*}
  {\Bigl(n_{(i_1,i_2)}^{(k)}\Bigr)_b}^a
  ={\Bigl(n_{(i_1,i_2)}^{(k)}\Bigr)_{\Psi^\prime(b)}}^{\Psi^\prime(a)}
  =2\sum_{\mu=0}^{k/2}\bar{S}_{\mu a}^{A_1}S_{\mu b}^{A_1}
  \chi_{(i_1,i_2)}^{A_2}\Bigl(-\frac{2\pi i}{k+3}(\mu+1,\mu+1)\Bigr)\quad.
\end{equation*}
  Remembering the definitions of $\mc{P}^{\prime\ast}$ in
  Theorem~\ref{thm:BranchingFormula} and of $\xi_\mu$ in
  equation~\eqref{eq:SMatrixCharacterRelation}, the argument of the character
  can be identified to be $\mc{P}^{\prime\ast}\xi_\mu$. By setting the
  index~$b$ to zero, Theorem~\ref{thm:BranchingFormula} implies
\begin{equation*}
  {b_{(i_1,i_2)}^\prime}^a
  =\lim_{k\to\infty}{\Bigl(n_{(i_1,i_2)}^{(k)}\Bigr)_0}^a
  =\lim_{k\to\infty}\sum_{\mu=0}^{k+1}\bar{S}_{\mu a}^{A_1}S_{\mu 0}^{A_1}
  \chi_{(i_1,i_2)}^{A_2}\bigl(\mc{P}^{\prime\ast}\xi_\mu\bigr)\quad.
\end{equation*}
  This equality holds because we are allowed to use the prefactor~$2$ to extend
  the range of~$\mu$ from $0\ldots,k/2$ to $0,\ldots,k+1$. Taking the
  considerations of the previous paragraph into account, we just proved that
  the NIM-rep for the twisted boundary conditions in the $A_2^{(1)}$ WZW model
  contains informations on both embeddings $A_1\hookrightarrow A_2$, with
  embedding index $x_e=1$ or $x_e^\prime=4$ respectively, at the same time.
  We leave it to the reader to write down the integral representation
  for branching coefficients of $A_1\hookrightarrow A_2$ with $x_e^\prime=4$.
  \smallskip

  After the detailed discussion of the $A_2$ case, we now want to comment
  on the $A_{2n}$ series for $n>1$. Numerical analysis indicates that a
  treatment similar to the one just presented leads to embeddings
  $B_n\hookrightarrow A_{2n}$, in addition to the embeddings
  $C_n\hookrightarrow A_{2n}$ which are proposed in~\cite{Quella:Other}.
  Following~\cite{Birke:1999ik}, the $A_{2n}^{(1)}$ NIM-rep labels are given by
  fractional symmetric weights $\alpha$ of~$A_{2n}$. To be more concrete, the
  Dynkin labels have to satisfy the relations $2\alpha_i\in\Natural_0$,
  $\alpha_i=\alpha_{2n+1-i}$ and $\sum_{i=0}^n\alpha_i\leq k/4$. Like before
  we assume the level~$k$ to be even. The map from weights of $B_n$ to the
  NIM-rep labels is then given by
\begin{equation*}
  \Psi^\prime\bigl(a_1,\cdots,a_n\bigr)
  =\frac{1}{4}\bigl(2a_{n-1},\cdots,2a_1,k-2a_1-2a_2-\cdots-2a_{n-1}-a_n,\cdots,2a_{n-1}\bigr)\quad.
\end{equation*}
  Again this map involves $k$ explicitly and is only well-defined for weights
  satisfying the relevant branching selection rule. We may use the projection
  $\mc{P}\bigl(i_1,\cdots,i_{2n}\bigr)
   =\bigl(i_1+i_{2n},i_2+i_{2n-1},\cdots,2(i_n+i_{n+1})\bigr)$
  to calculate the branching rules of $B_n\hookrightarrow A_{2n}$ according to
  Theorem~\ref{thm:RacahSpeiser} and compare them to NIM-rep calculations
  at $k\to\infty$ which have been performed using the algorithm proved
  in~\cite{Schweigert:Unpublished}. Taking our new identification
  of subalgebra weights with NIM-rep labels into account, full agreement has
  been observed. Up to now, however, we have no rigorous proof to support this
  observation. As a last remark, note that even in the case of $A_4$ our
  new identification requires a maximally embedded $B_2\cong C_2$ in contrast
  to the result in~\cite{Quella:Other}.

\section{\label{sc:Conclusions}Conclusions}

  In our paper we derived an explicit formula for the branching rules
  of embeddings of two semi-simple Lie algebras. Starting
  from this result, we gave an alternative proof for an algorithm which
  can be used to calculate branching rules. We have also been able to
  check some simple properties of branching coefficients explicitly and argued
  that our formula induces integral representations for them.
  In two examples, these integral representations have been derived
  explicitly.
  Finally, we discussed the relation of embeddings to NIM-reps of
  WZW models at infinite level. In particular we solved some puzzle which
  remained open in~\cite{Quella:Other} and found that one NIM-rep may
  contain informations about several embeddings at the same time by
  reinterpretation of NIM-rep labels. A possible continuation of general
  NIM-reps
  of the type~\eqref{eq:NIMRep} to finite values of~$k$ using a Verlinde-like
  formula~\eqref{eq:VerlindeFormula} might be of importance for a
  representation theoretic understanding of
  embeddings of quantum groups at roots of unity as it provides a
  natural analogue to the transition from tensor product to
  fusion coefficients (cmp.~\cite{Furlan:1990ce,GoodmanWenzl}).
  This last point has to be clarified in future work.
  Note that there has been some progress recently in understanding
  subgroups of quantum groups~\cite{Ocneanu1, Ocneanu2, Ocneanu3, Wassermann,
  Kirillov:2001ti}.
  \medskip

  Another approach to express the branching coefficients of semi-simple
  Lie algebras by using affine extensions of both Lie algebras
  at the same time would be to consider the grade
  zero part of the corresponding branching functions. A general expression for
  branching functions was found in~\cite{Hwang:1995yr}. However, it does not
  seem to provide a considerable simplification in our context.

\subsubsection*{Acknowledgements}

  The author likes to thank S.~Fredenhagen, J.~Fuchs, I.~Runkel,
  V.~Schomerus and Ch.~Schweigert for useful discussions and careful
  reading of the manuscript.
  In particular he is grateful to I.~Runkel and Ch.~Schweigert for
  the collaboration on~\cite{Schweigert:Unpublished}.
  This work was financially supported by the Studienstiftung des deutschen
  Volkes.

  \vspace{0.5cm}
  \noindent{\em Note added in proof:} After submission of this article two
  preprints~\cite{Petkova:2002yj,Gaberdiel:2002qa} have been published which
  discuss the relation of NIM-reps in $\mf{g}_k$ WZW models to certain
  subalgebras of~$\mf{g}$ and their affine extensions for finite values of the
  level~$k$.

\providecommand{\href}[2]{#2}\begingroup\raggedright\endgroup

\end{document}